\documentclass[aps,prb,reprint,superscriptaddress,floatfix]{revtex4-2}

\usepackage[T1]{fontenc}
\usepackage[utf8]{inputenc}
\usepackage{amsmath,amssymb,bm}
\usepackage{graphicx}
\usepackage{xcolor}
\usepackage[colorlinks=true,linkcolor=blue,citecolor=blue,urlcolor=blue]{hyperref}

\begin{document}
\raggedbottom

\title{Flat-band ferrimagnetism in the edge-centered honeycomb antiferromagnet: From harmonic flat magnon to broad continuum}

\author{Sicheng He}
\author{Zhixuan Zhao}
\author{Hanxiang Di}
\author{Shuaihua Zhang}
\author{Yanling Wu}
\author{Jun Li}
\email{ljcj007@ysu.edu.cn}
\affiliation{
	State Key Laboratory of Metastable Materials Science and Technology, Hebei Key Laboratory of Microstructural Material Physics, School of Science, Yanshan University, Qinhuangdao, 066004, China.}
\begin{abstract}
One of the most striking aspects of a flat band is not merely its lack of dispersion, but that it reduces the single-particle kinetic energy to a very small scale, allowing interaction effects to become the dominant energy scale in the low-energy physics. This picture has underpinned concepts such as flat-band ferromagnetism, flat-band superconductivity, and localized magnons. A more fundamental question, however, remains. Flat bands belong to the language of single-particle or linear-spin-wave physics; whether they survive as protected sharp excitations in a genuine many-body quantum magnet, or merely provide a skeleton of spectral weight left by the noninteracting picture, remains unclear. We study the spin-$1/2$ Heisenberg antiferromagnet on the edge-centered honeycomb lattice to examine this question. In contrast to flat bands in geometrically frustrated magnets such as kagome, this lattice has a $2:3$ sublattice imbalance that renders the intersublattice coupling matrix rectangular, whose null space gives rise to a flat band at the single-particle level. The same imbalance stabilizes a ferrimagnetic ground state through the Lieb--Mattis theorem and leads to an edge-site-localized flat magnon in linear spin-wave theory. Combining linear spin-wave theory, sign-problem-free stochastic series expansion quantum Monte Carlo, and stochastic analytic continuation, we find that quantum fluctuations transform this harmonic flat mode into a broad continuum whose centroid is about $1.55J$ and nearly dispersionless along the high-symmetry path. Sublattice imbalance thus offers an unfrustrated route that unifies flat bands and ferrimagnetism, and suggests that flat-band geometry organizes many-body spectral weight rather than giving rise to a protected quasiparticle.
\end{abstract}

\maketitle

\section{Introduction}
\label{sec:intro}

Flat bands, in which destructive interference quenches the kinetic energy, have become a major theme in correlated-electron and magnetic physics~\cite{leykam2018,derzhko2015}. Because their eigenstates can be selected as compact localized states (CLSs), interactions dominate the low-energy physics and can stabilize ordered phases that are absent in dispersive bands. Flat-band ferromagnetism in the Hubbard model~\cite{mielke1991,tasaki1992} and the localized magnons that produce magnetization jumps in frustrated quantum magnets~\cite{schulenburg2002} are prominent examples, and flat-band states have been observed directly in photonic Lieb lattices~\cite{mukherjee2015,vicencio2015}.

In magnetic systems, flat bands are most often discussed in the context of geometric frustration. In the kagome antiferromagnet, corner-sharing triangles produce an extensively degenerate classical ground-state manifold and a zero-energy spin-wave branch~\cite{harris1992,maksymenko2017}; in real materials, anisotropic interactions lift this branch to finite energy while leaving it weakly dispersive~\cite{matan2006}. A distinct, unfrustrated mechanism operates on bipartite lattices with unequal sublattice sizes~\cite{sutherland1986,inui1994}. There, the intersublattice hopping matrix is rectangular and therefore possesses a nontrivial null space, so that zero-energy states arise without odd cycles or competing interactions. Where such a flat band touches dispersive bands, its CLSs must be supplemented by extended states, a feature that reflects the real-space topology of the lattice~\cite{bergman2008,rhim2019}. The same sublattice imbalance has a second consequence for interacting spins and electrons: for the Hubbard model at half filling it enforces a ground state with nonzero total spin~\cite{lieb1989}, and for bipartite Heisenberg antiferromagnets it fixes the ground-state spin to the sublattice spin imbalance~\cite{lieb1962}. Such imbalance-driven ferrimagnetism has recently been observed with ultracold fermions in an optical Lieb lattice~\cite{lebrat2026}.

The edge-centered honeycomb lattice, obtained by placing an additional site on every bond of the honeycomb lattice, is a minimal two-dimensional realization of this mechanism with uniform nearest-neighbor coupling. Its unit cell contains two vertex sites and three edge-center sites, and its five-band single-particle spectrum hosts a flat band together with pseudospin-$1/2$ and pseudospin-1 Dirac points~\cite{shima1993,lan2012}. Localized flat-band states and noncontractible line states have been observed in photonic realizations~\cite{zhong2017,yan2020}. Magnetic studies of this geometry have addressed Dzyaloshinskii--Moriya-induced topological magnon bands, edge states, and thermal Hall responses~\cite{li2022}, as well as linear spin-wave thermodynamics in the presence of next-nearest-neighbor exchange~\cite{lima2024}. Here we examine the isotropic nearest-neighbor antiferromagnet in zero field.

Related lattice names have been used for different spin models. The kagome--honeycomb antiferromagnet of Adhikary \textit{et al.}~\cite{adhikary2021} has a six-site unit cell with three exchange couplings and exhibits a gapped hexagonal-singlet phase; all of its sites are threefold coordinated, and its bipartition is balanced ($3:3$). In the diamond-decorated honeycomb model~\cite{karlova2026}, conserved dimer spins reduce an eight-spin unit cell to an effective five-site description containing composite spin-0 or spin-1 sites and spin-$1/2$ monomers. Neither model coincides with the uniform five-site spin-$1/2$ Hamiltonian studied here.

For the Heisenberg model on such a lattice, the Lieb--Mattis ground multiplet carries an extensive uncompensated moment. Whether this moment is accompanied by long-range sublattice order in the thermodynamic limit, and how strongly quantum fluctuations reduce the sublattice moments, are separate questions that require an unbiased numerical treatment. The resulting state differs both from itinerant flat-band ferromagnetism and from a compensated N\'eel antiferromagnet. Quantum Monte Carlo (QMC) studies of mixed-spin Lieb-lattice and decorated-lattice Heisenberg models have shown that such ferrimagnetic states remain accessible without a sign problem~\cite{hirose2017,caci2023}.

The excitation spectrum raises a sharper question. Rank--nullity protects a single-particle zero mode, and the Lieb--Mattis theorem fixes only the ground-state spin; neither result constrains the finite-frequency magnons of the interacting system. In linear spin-wave theory (LSWT) the null vector reappears as a flat magnon branch, but magnon interactions can hybridize a mode with the multimagnon continuum whenever energy and momentum conservation allow it, broadening or even destroying the quasiparticle~\cite{zhitomirsky2013}. Answering this question requires dynamical information beyond spin-wave theory. Stochastic series expansion (SSE) QMC combined with stochastic analytic continuation (SAC) of imaginary-time correlations~\cite{shao2017,sandvik2016,shao2023} supplies such information and has resolved magnon modes, composite cluster excitations, and continua in checkerboard~\cite{xu2019,yan2021} and trimerized magnets~\cite{cheng2022,chang2024}, as well as in gapped spin chains~\cite{huang2021}. In particular, a broad and only weakly dispersive band near $\omega\approx J$ has been found in the ferrimagnetic Lieb-lattice Heisenberg model and attributed to destructive interference of composite excitations~\cite{chang2024}. These studies show that interaction effects reshape the higher-energy spectrum in ways that LSWT does not capture, and they motivate a direct test of the flat magnon in a lattice where its origin is fixed by sublattice imbalance alone.

In this paper we determine whether the localized excitation implied by the lattice geometry survives in the interacting edge-centered honeycomb antiferromagnet. We show that the $2:3$ sublattice imbalance simultaneously enforces the single-particle flat band, the ferrimagnetic ground-state spin, and a flat LSWT magnon at $\omega=2JS$, and that SSE simulations establish ferrimagnetic long-range order with substantially reduced sublattice moments. The dynamical calculations then show that the flat magnon does not survive as a sharp quasiparticle: its spectral weight forms a broad continuum whose centroid, near $1.55J$, is nearly momentum independent. Section~\ref{sec:model} introduces the lattice, the Hamiltonians, and the exact constraints. Section~\ref{sec:results} describes the Bloch and linear spin-wave theory (LSWT) analysis (Sec.~\ref{sec:lswt}), an exact Ising mapping that benchmarks the lattice (Sec.~\ref{sec:ising}), stochastic series expansion (SSE) quantum Monte Carlo results for the ground state (Sec.~\ref{sec:sse}), and the excitation spectrum obtained from exact spectral moments and stochastic analytic continuation (Secs.~\ref{sec:ed} and \ref{sec:sac}). Section~\ref{sec:extension} examines an unequal-exchange extension. The results are discussed in Sec.~\ref{sec:discussion} and summarized in Sec.~\ref{sec:conclusion}.

\begin{figure*}[tbp]
	\centering
	\includegraphics[width=0.9\textwidth]{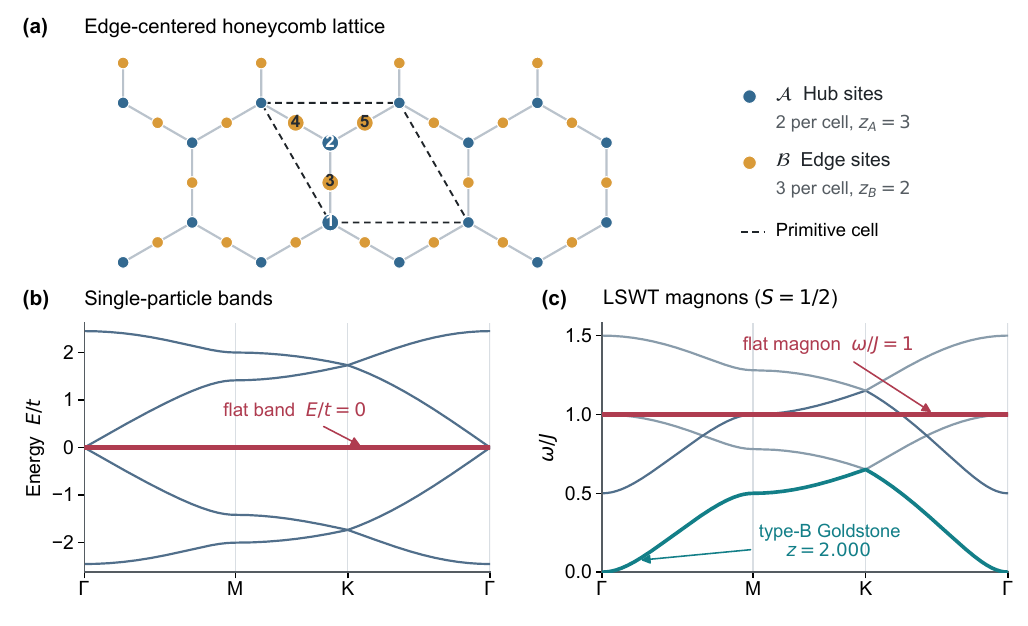}
	\caption{(a) Primitive cell and periodic nearest-neighbor graph; hub and edge sublattices are distinguished. (b) Single-particle bands along $\Gamma-M-K-\Gamma$, including the central flat band. (c) LSWT magnon bands for $S=1/2$; the flat branch at $\omega/J=1$ and the quadratic type-B Goldstone mode are emphasized.}
	\label{fig:structure-bands}
\end{figure*}
\section{Lattice, Hamiltonians, and exact constraints}
\label{sec:model}

The primitive cell comprises five sites [Fig.~\ref{fig:structure-bands}(a)]. We designate the two threefold-coordinated vertices of the underlying honeycomb lattice as the hub sublattice $\mathcal A$ and the three twofold-coordinated bond centers as the edge sublattice $\mathcal B$, so that
\begin{equation}
n_A=2,\qquad n_B=3,\qquad z_A=3,\qquad z_B=2.
\end{equation}

Because the primitive vectors are oblique, the first-neighbor shell was identified with the full Cartesian metric. It contains six bonds per unit cell at a distance of $0.2887$ in units of the cell coordinates. The coordination sequence is $(3,3,2,2,2)$, every bond links $\mathcal A$ to $\mathcal B$, and the two-coloring $\eta=(+1,+1,-1,-1,-1)$ shows that the graph is bipartite. Only this connectivity enters the model, and the primitive-vector length is set to unity in what follows.

Relative to a hub in cell $\mathbf R$, the adjacent edge sites are situated in the cells displaced by
\begin{equation}
\begin{array}{c|ccc}
 & C_1 & C_2 & C_3\\
A_1 & 0 & -\mathbf a_2 & -\mathbf a_1-\mathbf a_2\\
A_2 & 0 & 0 & 0
\end{array}.
\end{equation}
Here $A_{1,2}\in\mathcal A$ and $C_{1,2,3}\in\mathcal B$ label the sites within the primitive cell; an offset $\boldsymbol\delta$ connects the hub at $\mathbf R$ to the edge site at $\mathbf R+\boldsymbol\delta$ and contributes a phase $e^{i\mathbf k\cdot\boldsymbol\delta}$ to the Bloch Hamiltonian. The primitive vectors have equal length and enclose an angle of $120^\circ$. A periodic $L\times L$ cluster contains $N=5L^2$ sites and $N_b=6L^2$ bonds.

We examine the classical Ising model
\begin{equation}
H_{\rm I}=J\sum_{\langle ij\rangle}s_i s_j,
\qquad J>0,\quad s_i=\pm1,
\end{equation}
and the quantum Heisenberg model
\begin{equation}
H_{\rm H}=J\sum_{\langle ij\rangle}\mathbf S_i\cdot\mathbf S_j,
\qquad J>0,\quad S=\frac12.
\end{equation}
Both sums run over the six nearest-neighbor bonds per unit cell, and $J$ sets the energy scale. Same-sublattice exchange, anisotropy, Dzyaloshinskii--Moriya interactions, interlayer coupling, and external fields are absent. Because the lattice contains no odd cycles, the classical antiferromagnet is gauge equivalent to a ferromagnet. In the quantum case, a Marshall sign rotation renders all local matrix elements nonnegative, so that SSE operator strings carry positive weights~\cite{marshall1955,sandvik1999}.

The two sublattices contain $N_A=2L^2$ and $N_B=3L^2$ spins. The Lieb--Mattis theorem then fixes the ground-state total spin to
\begin{equation}
S_0=|N_B-N_A|S=\frac{L^2}{2},
\end{equation}
which corresponds to a moment of $S_0/N=1/10$ per site, or $1/5$ of the saturation value. The ground-state energy, the reduction of the sublattice moments by quantum fluctuations, the spin correlations, and the spin stiffness are not determined by this theorem; we compute them numerically in Sec.~\ref{sec:sse}.

\section{Results}
\label{sec:results}

\subsection{Bloch analysis and linear spin-wave theory}
\label{sec:lswt}

For uniform hopping $t$, the nearest-neighbor Bloch Hamiltonian takes the chiral form
\begin{gather}
H_0(\mathbf k)=t
\begin{pmatrix}
0&T(\mathbf k)\\
T^\dagger(\mathbf k)&0
\end{pmatrix},\\
T(\mathbf k)=
\begin{pmatrix}
1&e^{-ik_2}&e^{-i(k_1+k_2)}\\
1&1&1
\end{pmatrix}.
\end{gather}
With $k_i=\mathbf k\cdot\mathbf a_i$ and $f(\mathbf k)=1+e^{-ik_2}+e^{-i(k_1+k_2)}$, its five eigenvalues are
\begin{equation}
E_0(\mathbf k)=0,\qquad
E_{\pm,\pm}(\mathbf k)=\pm t\sqrt{3\pm|f(\mathbf k)|}.
\end{equation}
The zero-energy band follows without diagonalization: $T(\mathbf k)$ maps the three-dimensional edge subspace onto the two-dimensional hub subspace and therefore possesses a null vector at every momentum. Equivalently, the chiral operator $\Gamma=\operatorname{diag}(1,1,-1,-1,-1)$ anticommutes with $H_0$, $\{H_0,\Gamma\}=0$. The flat-band states reside entirely on sublattice $\mathcal B$. Numerical diagonalization of the periodic graph replicates this spectrum, including the flat band, to machine precision [Fig.~\ref{fig:structure-bands}(b)].

Alternating amplitudes on the six edge sites encircling a hexagon form a compact localized state (CLS), since their contributions cancel at every hub. Because the flat band touches the dispersive bands at $\Gamma$, these contractible CLSs do not span the full flat-band subspace on a torus, and noncontractible line states are required, in agreement with photonic observations~\cite{yan2020}. A counting argument makes this explicit. For $M=L^2$ unit cells, the vertex--edge incidence matrix has rank $2M-1$, so that the adjacency matrix has nullity $M+2$. The $M$ hexagon states obey one linear relation and span $M-1$ dimensions; two noncontractible line states complete the $(M+1)$-dimensional edge-site kernel, and one global vertex-site state supplies the remaining zero mode. The zero-energy degeneracy is therefore three at $\Gamma$ and one at all other momenta: $M$ states form the flat band proper, while the remaining two belong to the dispersive bands at the band-touching point.

The same rectangular block governs the harmonic magnon spectrum. Rotating the edge spins by $\pi$ relative to the hub spins and truncating the Holstein--Primakoff expansion at quadratic order, we obtain the bosonic dynamical matrix in one Nambu sector,
\begin{equation}
\mathcal D(\mathbf k)=JS
\begin{pmatrix}
3I_2&-C(\mathbf k)\\
C^\dagger(\mathbf k)&-2I_3
\end{pmatrix},
\end{equation}
where $C(\mathbf k)$ is a $2\times3$ intersublattice matrix whose singular values are determined by $3\pm|f(\mathbf k)|$. Diagonalization of $\mathcal D$ produces real frequencies on a dense momentum mesh. In contrast to the hopping problem, the local exchange field shifts the null sector of $C$ away from zero energy. With $g(\mathbf k)=|f(\mathbf k)|$, the five positive LSWT branches are
\begin{align}
\omega_{1,-}&=\frac{JS}{2}\left(\sqrt{13-4g}-1\right),\nonumber\\
\omega_{1,+}&=\frac{JS}{2}\left(\sqrt{13-4g}+1\right),\nonumber\\
\omega_{2,-}&=\frac{JS}{2}\left(\sqrt{13+4g}-1\right),\nonumber\\
\omega_{2,+}&=\frac{JS}{2}\left(\sqrt{13+4g}+1\right),\nonumber\\
\omega_{\rm flat}&=2JS,
\end{align}
in agreement with numerical diagonalization to machine precision. For $S=1/2$ the flat branch lies at $\omega_{\rm flat}/J=1$ [Fig.~\ref{fig:structure-bands}(c)]. Its eigenvector is confined to the three edge sites and cancels on both hubs; it thus has the same spatial structure as the single-particle CLS but appears at the finite frequency $2JS$ set by the exchange field of the ferrimagnetic background.

Near $\Gamma$, with $Q=k_1^2+k_1k_2+k_2^2$, the lowest branch behaves as
\begin{equation}
g=3-\frac{Q}{3}+O(k^4),\qquad
\omega_{1,-}=\frac{JS}{3}Q+O(k^4).
\end{equation}
For primitive vectors of length $a$ enclosing $120^\circ$, $|\mathbf q|^2=4Q/(3a^2)$, and hence
\begin{equation}
\omega_{1,-}=\frac{JSa^2}{4}|\mathbf q|^2+O(q^4).
\end{equation}
A numerical fit to the lowest branch yields $z=2.000$ and, for $a=1$ and $S=1/2$, a prefactor of $0.125J$, in agreement with the analytic value $J/8$. The spectrum consequently contains a single quadratic type-B Goldstone mode instead of the two linearly dispersing modes of a collinear antiferromagnet: the finite ground-state magnetization renders the two broken SU(2) generators canonically conjugate~\cite{watanabe2012}. The branch of opposite precession is gapped, with energy $JS=0.5J$ at $\Gamma$.

The structure of the harmonic spectrum follows from the singular-value decomposition of $C(\mathbf k)$, whose singular values satisfy
\begin{equation}
\sigma_{1,2}^2=3\pm g(\mathbf k).
\end{equation}
For each nonzero singular value $\sigma$, the dynamical matrix reduces to the $2\times2$ block
\begin{equation}
\begin{pmatrix}
3&-\sigma\\
\sigma&-2
\end{pmatrix},
\end{equation}
with eigenvalues
\begin{equation}
\lambda_{\pm}=\frac{1\pm\sqrt{25-4\sigma^2}}{2}.
\end{equation}
The two singular values, together with the conjugate Nambu sector, produce the four dispersive branches given above. For generic momenta, $C$ has a one-dimensional right null space, and any $v_0\in\ker C$ satisfies
\begin{equation}
\mathcal D(\mathbf k)
\begin{pmatrix}0\\v_0\end{pmatrix}
=-2JS
\begin{pmatrix}0\\v_0\end{pmatrix}.
\end{equation}
The conjugate sector supplies the corresponding positive frequency $2JS$. This branch is momentum independent because all edge sites experience the same local exchange field, $z_BJS=2JS$.

\subsection{Decoration mapping and Wolff cluster simulations}
\label{sec:ising}

Tracing out the edge spin on each decorated bond gives
\begin{equation}
\sum_{s=\pm1}e^{-K s(s_1+s_2)}
=2\cosh[K(s_1+s_2)],\qquad K=J/T.
\end{equation}
The ratio of the Boltzmann weights for parallel and antiparallel hub spins is $\cosh(2K)$, so that the decoration transformation~\cite{fisher1959} yields the effective honeycomb coupling
\begin{equation}
K_{\rm eff}=\frac12\ln[\cosh(2K)].
\end{equation}
Combined with the exact critical condition of the honeycomb Ising model, $\cosh(2K_{{\rm eff},c})=2$, i.e., $K_{{\rm eff},c}=\tfrac12\ln(2+\sqrt3)$~\cite{husimi1950,houtappel1950}, this yields
\begin{equation}
\frac{T_c^{\rm Ising}}{J}
=\frac{2}{\operatorname{arcosh}(2+\sqrt3)}
=1.0041913\ldots.
\label{eq:tc}
\end{equation}
In terms of the initial variables, the ordered phase is ferrimagnetic: at $T=0$ the two hub spins and the three edge spins of each cell are antiparallel, so that the staggered moment is saturated and the net moment is $|M|/N=1/5$.

To test this result directly on the five-site lattice, we conducted Wolff cluster simulations of the original model. The gauge-transformed variables $\tau_i=\eta_i s_i$ form a ferromagnetic Ising model, to which the Wolff algorithm applies with bond-activation probability $1-e^{-2J/T}$~\cite{wolff1989}; one Monte Carlo step consists of one cluster construction and flip. We simulated $L=8,12,16,24,32,48,64$ at $T/J=0.996,1.000,1.004,1.008,1.012$, with 5000 equilibration steps and 64 bins of 2000 measurements each, and computed the staggered Binder cumulant
\begin{equation}
U_4=1-\frac{\langle M_s^4\rangle}{3\langle M_s^2\rangle^2},
\qquad M_s=\sum_i\eta_i s_i.
\end{equation}
Crossing temperatures were obtained by linear interpolation between adjacent temperatures. Statistical errors were estimated by a delete-one-block jackknife with block sizes of 1, 2, 4, and 8 bins, taking the largest value, and were propagated to the crossings.

\begin{figure}[tbp]
\centering
\includegraphics[width=\columnwidth]{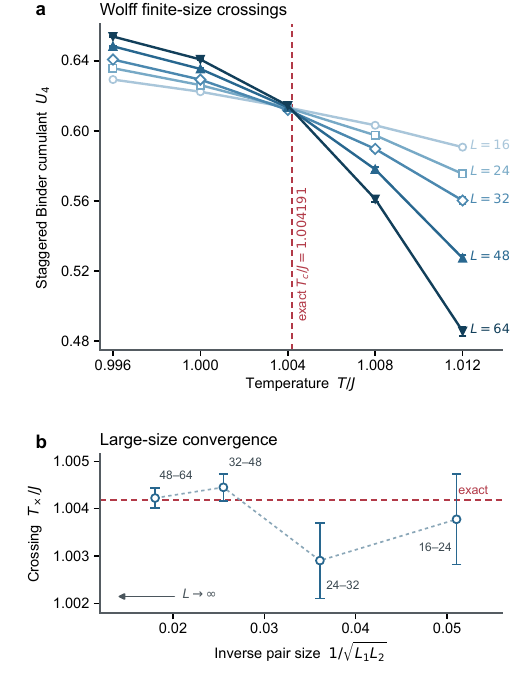}
\caption{(a) Staggered Binder cumulant for the 35 simulated parameter points; the vertical line denotes the exact critical temperature of Eq.~(\ref{eq:tc}). (b) Crossing temperatures of pairs of system sizes versus inverse geometric-mean size. Error bars indicate standard errors from 1/2/4/8-bin reblocking.}
\label{fig:ising}
\end{figure}

The Binder cumulants are presented in Fig.~\ref{fig:ising}. The crossing temperatures approach the exact result with increasing system size, $T_c/J=1.004451(285)$ for $L=32,48$ and $1.004225(206)$ for $L=48,64$; the latter agrees with Eq.~(\ref{eq:tc}) within one standard error. We further interpolated $|M_s|$ and the connected staggered susceptibility to the exact $T_c$. Power-law fits for $L\geq8$ yield $\beta/\nu=0.12549$ (95\% parametric-bootstrap interval $0.12378$--$0.12717$) and $\gamma/\nu=1.7584$ ($1.7467$--$1.7698$); limiting the fits to $L\geq16$ gives $0.12570$ ($0.12284$--$0.12863$) and $1.7502$ ($1.7308$--$1.7700$). All intervals contain the exact two-dimensional Ising values $1/8$ and $7/4$~\cite{yang1952,wu1976}, and the Binder cumulants collapse with $\nu=1$. The transition on the decorated lattice thus belongs to the two-dimensional Ising universality class, as required by the decoration mapping.

\subsection{Ground-state properties from stochastic series expansion}
\label{sec:sse}

\begin{figure*}[tbp]
	\centering
	\includegraphics[width=0.9\textwidth]{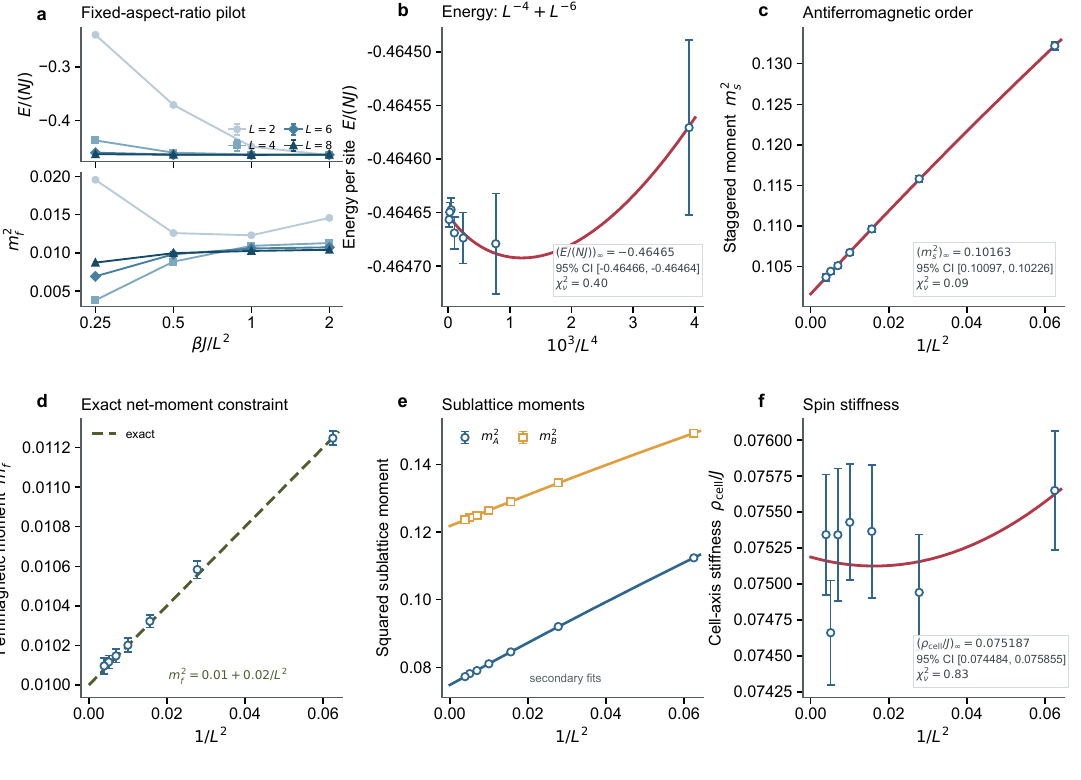}
	\caption{(a) Energy and squared ferrimagnetic moment versus $\beta J/L^2$ from the preliminary scan for $L=2,4,6,8$. (b) Energy per site with a fit to $e_0+a_4L^{-4}+a_6L^{-6}$. (c) Squared staggered moment with a fit to $m_{s,\infty}^2+b_2L^{-2}+b_4L^{-4}$. (d) Squared ferrimagnetic moment compared with the finite-size Lieb--Mattis value $0.01+0.02/L^2$. (e) Squared hub and edge sublattice moments with direct fits. (f) Spin stiffness in cell-axis units with a fit to $\rho_{{\rm cell},\infty}+c_2L^{-2}+c_4L^{-4}$. Data in (b)--(f) combine two independent Markov chains (128 bins per size); error bars are the largest of the unblocked and chain-resolved 2/4/8-bin standard errors. Intervals reported in the panels are 95\% parametric-bootstrap statistical intervals.}
	\label{fig:sse}
\end{figure*}
We simulated the Heisenberg model with the SSE method, which samples the expansion
\begin{equation}
Z=\sum_{\alpha}\sum_{n=0}^{\infty}
\frac{\beta^n}{n!}
\langle\alpha|(-H)^n|\alpha\rangle,
\end{equation}
applying diagonal updates and operator-loop updates~\cite{sandvik1999}. The implementation was validated against established results for the square lattice~\cite{sandvik1997} and against exact diagonalization on the square-octagon lattice, the lattice of octagraphene~\cite{li2020}.

We measured the energy per site, the squared uniform (ferrimagnetic) moment
\begin{equation}
m_f^2=\frac{3}{N^2}\left\langle(M_A^z+M_B^z)^2\right\rangle,
\end{equation}
the squared staggered moment
\begin{equation}
m_s^2=\frac{3}{N^2}\left\langle(M_A^z-M_B^z)^2\right\rangle,
\end{equation}
the squared sublattice moments $3\langle(M_{A,B}^z)^2\rangle/N_{A,B}^2$ and their cross correlation, the uniform susceptibility, and the spin stiffness from winding-number fluctuations. Here $M_A^z=\sum_{i\in\mathcal A}S_i^z$ and $M_B^z=\sum_{i\in\mathcal B}S_i^z$. For an SU(2)-symmetric model, averaging over the two winding directions yields the standard prefactor $3/(4\beta)$~\cite{sandvik1997}; in primitive-cell coordinates, the helicity modulus is
\begin{equation}
\rho_{\rm cell}=\frac{3}{4\beta}\left(\langle W_1^2\rangle+\langle W_2^2\rangle\right).
\end{equation}
We adopt this cell-axis convention throughout; for equal-length primitive vectors enclosing $120^\circ$, the Cartesian stiffness is $\rho_{\rm Cartesian}=(\sqrt3/2)\rho_{\rm cell}$. For a finite cluster whose ground multiplet is sampled in zero field, SU(2) symmetry implies
\begin{equation}
m_{f,0}^2(L)=\frac{S_0(S_0+1)}{N^2}
=\frac{1}{100}+\frac{1}{50L^2}.
\end{equation}

Because the quadratic Goldstone mode closes the finite-size gap as $L^{-2}$, the inverse temperature must scale as $\beta\propto L^2$ to approach the ground state uniformly in system size. A preliminary scan over $\beta J/L^2=0.25$--2 for $L=2$--8 [Fig.~\ref{fig:sse}(a)] shows that, for $L\geq4$, the energy changes only slightly between the two largest ratios and the ferrimagnetic moment approaches its zero-temperature Lieb--Mattis value at $\beta J/L^2=2$ (Table~\ref{tab:pilot}). The $L=2$ cluster remains temperature dependent between $\beta J/L^2=1$ and 2 and was not included in the subsequent analysis. The main simulations were therefore performed at $\beta J=2L^2$ for $L=4,6,\ldots,16$, with two independent Markov chains per size, $10^4$ equilibration sweeps, and 64 bins of $10^3$ sweeps per chain. All 14 simulations passed stationarity, inter-chain consistency, and reblocking tests and were included in the analysis. Error bars are the largest of the unblocked and the chain-resolved 2/4/8-bin standard errors of the 128 pooled bins.

\begin{table}[tbp]
\caption{Preliminary SSE results at $\beta J=2L^2$ (32 bins of 500 sweeps). The squared ferrimagnetic moment is compared with the finite-size Lieb--Mattis value $S_0(S_0+1)/N^2$.}
\label{tab:pilot}
\begin{ruledtabular}
\begin{tabular}{ccccc}
$L$ & $E/(NJ)$ & $m_f^2$ & $S_0(S_0+1)/N^2$ & $\rho_{\rm cell}/J$ \\
\hline
4 & $-0.46431(21)$ & $0.01128(7)$ & $0.011250$ & $0.0755(10)$ \\
6 & $-0.46467(9)$ & $0.01072(9)$ & $0.010556$ & $0.0741(9)$ \\
8 & $-0.46467(7)$ & $0.01040(10)$ & $0.010313$ & $0.0753(10)$ \\
\end{tabular}
\end{ruledtabular}
\end{table}

We extrapolated to the thermodynamic limit using
\begin{align}
\frac{E(L)}{N}&=e_0+\frac{a_4}{L^4}+\frac{a_6}{L^6},\nonumber\\
m_s^2(L)&=m_{s,\infty}^2+\frac{b_2}{L^2}+\frac{b_4}{L^4},
\end{align}
and
\begin{equation}
\rho_{\rm cell}(L)=\rho_{{\rm cell},\infty}+\frac{c_2}{L^2}+\frac{c_4}{L^4}.
\end{equation}
The $L^{-4}$ correction to the energy follows from the Goldstone contribution with $d+z=4$; the forms for $m_s^2$ and $\rho_{\rm cell}$ are even-power finite-size ansatzes. The squared sublattice moments and their cross correlation were fitted in the identical way. All fits were performed by weighted least squares, and confidence intervals were derived from 10000 parametric bootstrap resamples.

\begin{table*}[tbp]
\caption{Thermodynamic-limit estimates from fits over $L=4$--16. Intervals are 95\% parametric-bootstrap statistical intervals for the specified finite-size forms. The last column shows the range of intercepts obtained with $L_{\min}=6$ and 8.}
\label{tab:fss}
\begin{ruledtabular}
\begin{tabular}{lcccc}
Quantity & Estimate & 95\% interval & $\chi^2/{\rm dof}$ & $L_{\min}=6,8$ \\
\hline
$E_0/(NJ)$ & $-0.464652$ & $[-0.464663,-0.464641]$ & $0.40$ & $[-0.464653,-0.464651]$ \\
$m_{s,\infty}^2$ & $0.10163$ & $[0.10097,0.10226]$ & $0.09$ & $[0.10163,0.10198]$ \\
$\rho_{{\rm cell},\infty}/J$ & $0.07519$ & $[0.07448,0.07586]$ & $0.83$ & $[0.07463,0.07519]$ \\
$m_A^2$ & $0.07478$ & $[0.07427,0.07527]$ & $0.13$ & -- \\
$m_B^2$ & $0.12181$ & $[0.12104,0.12256]$ & $0.08$ & -- \\
$\langle\mathbf M_A\!\cdot\!\mathbf M_B\rangle/(N_AN_B)$ & $-0.09544$ & $[-0.09607,-0.09485]$ & $0.10$ & -- \\
\end{tabular}
\end{ruledtabular}
\end{table*}

The extrapolated ground-state quantities are summarized in Table~\ref{tab:fss} and Fig.~\ref{fig:sse}. The ferrimagnetic moment follows the exact finite-size value $m_{f,0}^2=0.01+0.02/L^2$ with a maximum standardized deviation of $0.62$, as required by the Lieb--Mattis theorem. The quantities renormalized by quantum fluctuations are the energy, the staggered and sublattice moments, and the stiffness. Fits over $L=4$--16 give $E_0/(NJ)=-0.464652$, $m_{s,\infty}^2=0.10163$, and $\rho_{{\rm cell},\infty}/J=0.07519$, with $\chi^2/{\rm dof}\leq0.83$; increasing $L_{\min}$ to 6 or 8 shifts the intercepts by less than their statistical uncertainties. The finite staggered moment and stiffness establish ferrimagnetic long-range order in the ground state. Direct fits of the sublattice moments yield $m_A^2=0.07478$ and $m_B^2=0.12181$, together with a negative cross correlation. Assuming collinear antiparallel order and factorization of the cross term in the thermodynamic limit, the staggered fit alternatively implies $m_A=0.2735$ ($0.2722$--$0.2747$) and $m_B=0.34899$ ($0.34813$--$0.34981$).

Exact diagonalization of the $L=2$ ($N=20$) cluster offers an independent verification. The lowest energies in the $S^z=0,1,2$ sectors coincide to within $1.4\times10^{-14}J$, whereas the lowest $S^z=3$ state lies $0.97038J$ higher, and $\langle\mathbf S_{\rm tot}^2\rangle=6.000$ identifies the $S_0=2$ multiplet. The ground-state energy is $E_0/(NJ)=-0.46656$, the ferrimagnetic moment $m_f^2=0.015000$ equals the finite-cluster Lieb--Mattis value, and the staggered moment is $m_s^2=0.20989$.

\subsection{Spectral moments from exact diagonalization}
\label{sec:ed}

If the harmonic flat magnon survived as an exact eigenmode of the interacting Hamiltonian, a spin operator based on the single-particle null vector would create an excitation of a single energy. We tested this on the $N=20$ cluster by applying a CLS spin operator to the highest-weight ground state $|0\rangle$ with $S^z=S_0=2$. The operator acts on the six edge sites surrounding one hexagon with alternating coefficients $c_i=\pm1/\sqrt6$ and vanishes elsewhere; the coefficients satisfy $\sum_i c_i^2=1$ and $Ac=0$, where $A$ is the single-particle adjacency matrix. For $\alpha=-,z,+$, we defined $O_c^\alpha=\sum_i c_iS_i^\alpha$, projected out the ground multiplet, and normalized the resulting state $|\phi_\alpha\rangle$. Its first two spectral moments are
\begin{gather}
\bar\omega_\alpha=\langle\phi_\alpha|(H-E_0)|\phi_\alpha\rangle,\\
\sigma_\alpha^2=\|(H-E_0-\bar\omega_\alpha)|\phi_\alpha\rangle\|^2.
\end{gather}
A nonzero variance $\sigma_\alpha^2$ excludes an excitation of a single, possibly degenerate, energy; the frequency distribution of the weight is obtained in Sec.~\ref{sec:sac}.

The fixed-$S^z$ sectors have dimensions 167960, 125970, and 77520 for the lowering, longitudinal, and raising channels, respectively. The ground state was calculated with a sparse Lanczos solver at a tolerance of $10^{-12}$, and the moments were evaluated by direct application of the Hamiltonian, without broadening or analytic continuation. We also constructed translated CLS operators at cell momenta $(\pi,0)$, $(0,\pi)$, and $(\pi,\pi)$; their $\Gamma$-point component vanishes identically, so this momentum is not accessible with the current operator.

The ground-state energy, $E_0/J=-9.33115$, agrees with the result of Sec.~\ref{sec:sse}. In all three channels the CLS state has a finite energy variance. The lowering, longitudinal, and raising channels have mean excitation energies of $1.35504J$, $1.87471J$, and $2.58255J$, standard deviations of $0.47800J$, $0.36463J$, and $0.34238J$, and integrated inelastic weights of $0.71384$, $0.12586$, and $0.04493$ for the normalization defined above, respectively. Within each channel, the three momentum-resolved operators yield identical moments to numerical precision and retain a finite variance, so the energy spread cannot be attributed to a superposition of different momenta of a single dispersive branch. On a finite cluster, the variance measures the width of a discrete spectrum rather than a decay rate. The asymmetry among the channels reflects the use of the highest-weight state, whereas the zero-field SSE samples the complete ground multiplet; a comparison with the SSE spectra therefore necessitates averaging over the multiplet. The CLS operator thus does not create a single-energy excitation, and its spectral distribution on larger lattices is determined below.

\subsection{Frequency-resolved spectra from stochastic analytic continuation}
\label{sec:sac}

To resolve the spectral distribution on larger lattices, we calculated the imaginary-time correlation functions
\begin{equation}
C_\lambda(\mathbf q,\tau)=\langle O_\lambda(\mathbf q,\tau)O_\lambda^\dagger(\mathbf q,0)\rangle
\end{equation}
for the total, staggered, hub-only, edge-only, and CLS (edge null-vector) form factors along the $\Gamma$--M--K--$\Gamma$ path, which contains 7 momenta for $L=6$ and 13 for $L=12$. Both sizes were simulated at $\beta J=2L^2$ with two independent Markov chains, 5000 equilibration sweeps, and 32 bins of 200 sweeps per chain, giving 64 bins per size. The correlation functions were measured every 20 sweeps on a grid with $\Delta\tau J=0.25$, supplemented by sparse points at large $\tau$.

Spectral functions were obtained by SAC~\cite{shao2017,shao2023}, using the bootstrap covariance matrix in its eigenbasis. The primary continuations adopt an unconstrained distribution of 300 $\delta$ functions, 1000 bootstrap samples, 500 annealing steps, and 50 sampling bins of 5000 steps each. To evaluate the sensitivity to the continuation, we repeated the CLS analysis with an independent random seed, with a parametrization containing a dominant $\delta$ function, and with data thinned by a factor of two in both bins and imaginary-time points. Because the total spin is conserved, the total-spin response at $\Gamma$ is purely elastic and was excluded from the continuation; the same applies to the $\Gamma$-point edge channel whenever its normalized endpoint exceeds $0.98$. Applied to isolated Heisenberg dimers at $\beta J=8$, the procedure places the spectral maximum at $0.925J$ and the first moment at $1.029J$, compared with the exact gap $J$.

\begin{figure*}[tbp]
\centering
\includegraphics[width=\textwidth]{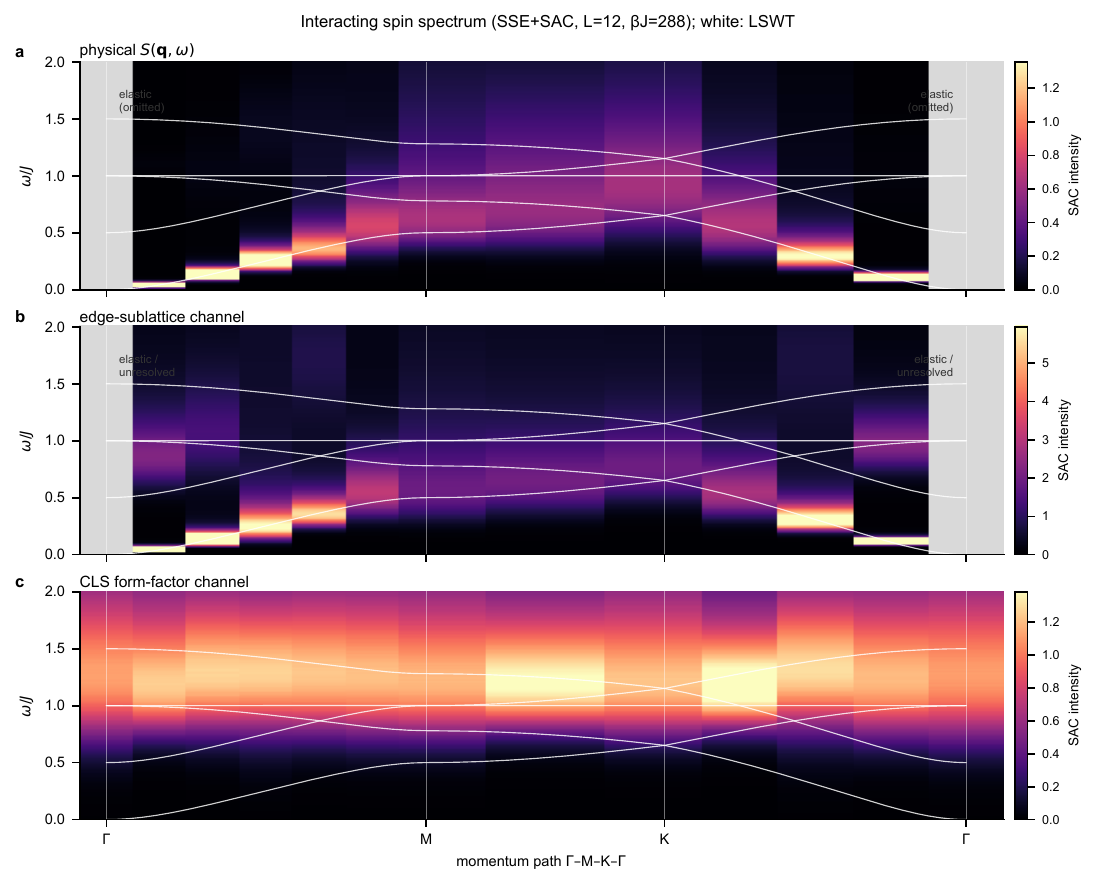}
\caption{Dynamical spectra for $L=12$ and $\beta J=288$ obtained by stochastic analytic continuation of 64 bins: (a) total-spin, (b) edge-sublattice, and (c) CLS form-factor channel along $\Gamma$--M--K--$\Gamma$. White lines represent the five LSWT branches. Each panel has its own intensity scale. Gray regions at the path endpoints denote elastic or unresolved $\Gamma$-point channels.}
\label{fig:sac-spectrum}
\end{figure*}

Figure~\ref{fig:sac-spectrum} shows the spectra for $L=12$. The total-spin channel is dominated by the low-energy dispersive Goldstone response, while the edge-projected channel contains both this response and finite-frequency weight. The CLS form factor suppresses the Goldstone contribution and isolates a broad band of spectral weight above the LSWT flat-band energy $\omega/J=1$. This band spans an energy interval of order $J$ and is therefore not a $\delta$-like branch, consistent with the finite spectral variance obtained for $N=20$.

For the primary continuation at $L=12$, the position of the CLS maximum varies between $1.165J$ and $1.335J$ along the 13-point path, whereas the spectral centroid varies only between $1.517J$ and $1.566J$, i.e., over a bandwidth of $0.050J$; the integrated CLS weight varies by less than 1\%. For $L=6$ the centroid bandwidth is $0.072J$, and the mean centroid decreases from $1.586J$ for $L=6$ to $1.548J$ for $L=12$.

The centroid is also robust against changes of the continuation procedure, whereas the peak position is not. For $L=12$, the independent seed, the dominant-$\delta$ parametrization, and the thinned data give mean centroids of $1.548J$, $1.553J$, and $1.545J$ with momentum variations of $0.054J$, $0.059J$, and $0.042J$, respectively, while the corresponding variations of the maximum position are $0.100J$, $0.640J$, and $0.220J$. The CLS-projected spectral weight therefore forms a weakly dispersive continuum centered near $1.55J$, above the harmonic flat-band energy, without the stable narrow peak anticipated for a protected flat magnon.

\subsection{Unequal-exchange extension}
\label{sec:extension}

Making the two hubs inequivalent connects the uniform model to an exactly solvable limit of decoupled clusters. We preserve the five-site connectivity but assign the exchange $J_A$ to the three bonds of hub 1 and $J_B$ to the three bonds of hub 2 in each cell. Both hubs belong to sublattice $\mathcal A$; the labels of the couplings do not refer to the sublattices. We set $J_A=1$ and define $r=J_B/J_A>0$. Separate counts of the two operator types in the SSE expansion give the bond energies $C_t=1/4-\langle n_t\rangle/(\beta N_{b,t}J_t)$, with $N_{b,A}=N_{b,B}=3L^2$. The heat capacity per spin is $c=(\langle n^2\rangle-\langle n\rangle^2-\langle n\rangle)/N$, with $n=n_A+n_B$. Because this estimator is nonlinear, its uncertainty was obtained from a delete-one-bin jackknife over the first and second moments; for linear estimators we employed the standard error over bins.

The method was benchmarked on isolated dimers of both coupling types, whose thermodynamics is known exactly, at $L=2$, $r=0.1$ and 1, and $T/J_A=0.05$, 0.3, and 2 (two Markov chains, 64 bins of 1000 sweeps); agreement was required within $|\Delta|\leq5\,{\rm SE}+10^{-4}$. The lattice simulations include $L=4$ and 8, $r=0.02$, 0.1, and 1, and $T/J_A=0.01,0.03,0.1,0.3,1,3$, with two Markov chains and 32 bins of 500 sweeps per point. These data illustrate temperature trends at fixed system size and are not extrapolated to the thermodynamic limit.

\paragraph{Decoupled-star limit.}
\label{app:star}
For $J_B=0$, the lattice decomposes into four-spin stars and free hub-2 spins. The star Hamiltonian is
\begin{equation}
H_\star=J_A\mathbf s_0\cdot\mathbf R,\qquad
\mathbf R=\mathbf s_1+\mathbf s_2+\mathbf s_3.
\end{equation}
Coupling the central spin $1/2$ to $R=1/2$ (which occurs twice) and $R=3/2$ produces the star energies $E/J_A=(-5/4,-3/4,1/4,3/4)$, with degeneracies $(3,2,6,5)$ and total spins $(1,0,1,2)$. With $x=J_A/T$ and $q=e^{-x/2}$, we define
\begin{equation}
D=3+2q+6q^3+5q^4.
\end{equation}
The partition function per cell is $Z_{\rm cell}=2e^{5x/4}D$, where the factor 2 allows for the free spin. With
\begin{equation}
a=\frac{q+9q^3+10q^4}{D},
\end{equation}
\begin{equation}
b=\frac{q/2+27q^3/2+20q^4}{D},
\end{equation}
the energy, heat capacity, entropy, and uniform susceptibility per spin are
\begin{equation}
e=\frac{J_A}{5}\left(-\frac54+a\right),
\end{equation}
\begin{equation}
c=\frac{x^2}{5}(b-a^2),\qquad s=\frac{\ln(2D)+xa}{5},
\end{equation}
\begin{equation}
T\chi=\frac15\left[\frac14+\frac{2+4q^3+10q^4}{D}\right],
\end{equation}
where $\chi=\langle(S^z_{\rm tot})^2\rangle/(NT)$ in zero field and $N=5N_c$. The strong-bond correlation is $5e/(3J_A)$. For $T\to0$, $e\to-J_A/4$, $s\to\ln6/5$, and $T\chi\to11/60$; at high temperature, $e=-9J_A^2/(80T)+O(J_A^3/T^2)$, $c=9J_A^2/(80T^2)+O(x^3)$, and $T\chi=1/4-3x/40+O(x^2)$.

The leading correction in $J_B$ can be derived without assuming that the perturbation commutes with $H_\star$. Since the squared total magnetization commutes with the unperturbed Hamiltonian, cyclicity of the trace reduces the first-order derivative to a covariance. With
\begin{equation}
F(T)=\langle\mathbf T_\star\cdot\mathbf s_e\rangle_0
=\frac{5/2+2q^3+15q^4/2}{D},
\end{equation}
where $\mathbf T_\star$ is the total star spin, each weak bond contributes $F/6$ to $\langle(S^z_{\rm tot})^2\,\mathbf s_e\cdot\mathbf s_B\rangle_0$. Summing over the $3N_c$ weak bonds and dividing by $5N_c$ gives, at fixed $T>0$,
\begin{equation}
T\chi(T,J_B)=T\chi(T,0)-\frac{J_BF(T)}{10T}+O_T(J_B^2).
\end{equation}
Since $F\to5/6$ when the star excitations freeze out, the leading correction is $-J_B/(12T)$; at high temperature $F\to3/4$, which recovers $T\chi=1/4-3(J_A+J_B)/(40T)$ to first order in exchange over temperature. The closed-form expressions agree with a direct trace over the 32-dimensional Hilbert space of a star and a free spin to machine precision, and a central finite difference recovers the derivative to $1.2\times10^{-7}$.

The order of limits is crucial. For fixed $J_B>0$, a connected cluster of $N_c$ cells has the ground-state spin $S_0=N_c/2$, so that its zero-temperature multiplet gives $s_0=\ln(N_c+1)/(5N_c)$ and $T\chi=(N_c+2)/60$. Neither coincides with the intermediate-temperature values of independently fluctuating star triplets and free spins.

All 12 benchmark simulations agree with the exact dimer results in all 60 comparisons. At the lowest temperatures the statistical resolution becomes limiting: a pooled estimate at $r=0.1$, $T/J_A=0.05$ gives $c=0.279\pm0.024$, about three standard errors above the exact value $0.2054$, and at $r=1$, $T/J_A=0.05$ no excitations were sampled, giving $T\chi=0\pm0$ compared with the exact value $2.1\times10^{-9}$.

For $r=0.02$, both system sizes and both Markov chains show strong-bond correlations forming at higher temperatures than weak-bond correlations (Table~\ref{tab:thermal-pilot}). For $L=8$, the strong-bond correlation is approximately $-0.205$ at $T/J_A=1$ and $-0.391$ at 0.3 and remains close to the decoupled-star value $-5/12$ between 0.1 and 0.01, while the magnitude of the weak-bond correlation rises from approximately 0.024 to 0.169.

\begin{table}[tbp]
\caption{Bond correlations and susceptibility for $L=8$ and $r=0.02$, presented separately for the two Markov chains. Numbers in parentheses are bin-based standard errors in the last digits. $C_A$ and $C_B$ are bond averages, and $\chi$ is given per spin.}
\label{tab:thermal-pilot}
\begin{ruledtabular}
\begin{tabular}{ccccc}
$T/J_A$ & Chain & $C_A$ & $C_B$ & $T\chi$\\
\hline
0.10 & 1 & $-0.41494(33)$ & $-0.02420(91)$ & $0.1663(19)$\\
0.10 & 2 & $-0.41648(44)$ & $-0.02395(69)$ & $0.1661(20)$\\
0.01 & 1 & $-0.41678(14)$ & $-0.16869(50)$ & $0.1022(10)$\\
0.01 & 2 & $-0.41652(14)$ & $-0.17008(47)$ & $0.1024(12)$\\
\end{tabular}
\end{ruledtabular}
\end{table}

The low-temperature susceptibility is strongly size dependent. At $T/J_A=0.01$, $T\chi$ is approximately 0.291 for $L=4$ and 0.925 for $L=8$ at $r=0.1$, and 0.302 and 1.096, respectively, at $r=1$, reflecting finite-size gaps. At $T/J_A\leq0.03$ the statistical errors of the heat capacity are comparable to its magnitude, so that no low-temperature peak structure can be discerned.

\section{Discussion}
\label{sec:discussion}

The results identify a common geometric origin for three phenomena that are typically treated separately. The $2:3$ sublattice imbalance renders the intersublattice coupling block rectangular. For the hopping problem, rank--nullity then guarantees an edge-supported zero mode at every momentum; for the spin problem, the same imbalance fixes an extensive ground-state spin through the Lieb--Mattis theorem; and in LSWT the null vector reappears as an edge-supported mode, shifted to the finite frequency $2JS$ by the local exchange field. The flat band, the ferrimagnetic moment, and the harmonic flat magnon are thus outcomes of a single lattice constraint, although each requires a separate argument.

The interacting calculations determine how far this relationship extends. The SSE ground-state results follow the exact total-spin constraint and yield a finite staggered moment and spin stiffness, establishing a quantum ferrimagnet with a quadratic type-B Goldstone mode. The dynamical results show, in contrast, that the harmonic flat branch does not survive as a sharp many-body excitation. The CLS operator has a finite spectral variance already on the 20-site cluster, and on larger lattices the corresponding spectral weight is distributed over an energy range of order $J$. The weak dispersion of its centroid reflects the destructive interference inherited from the lattice geometry, whereas the continuation-dependent maximum and the large spectral width preclude a quasiparticle interpretation. Sublattice imbalance thus organizes the spectral weight but does not prevent its interaction-induced hybridization and broadening.

The absence of a sharp flat magnon has a transparent origin. In the fully polarized phase the one-magnon sector is exactly decoupled from all others, so that CLSs are exact eigenstates and the flat band is protected~\cite{schulenburg2002}. The zero-field ferrimagnet offers no such protection. Its ground state is strongly entangled: quantum fluctuations reduce the sublattice moments to $m_A\approx0.27$ and $m_B\approx0.35$, well below the classical value $S=1/2$ (Table~\ref{tab:fss}), and magnon interactions couple the harmonic modes to multimagnon states. Kinematically, the flat branch at $\omega=J$ for $S=1/2$ lies inside the energy window $0\le\omega\le(\sqrt{13}-1)J/2\approx1.30J$ spanned by pairs of Goldstone-branch magnons, so that its decay into multimagnon states is not kinematically forbidden~\cite{zhitomirsky2013}. Destructive interference still suppresses the motion of the localized amplitude, which explains why the centroid of the CLS-projected spectrum remains almost momentum independent, but it no longer restricts the weight to a single energy. The upward shift of the centroid from the harmonic value $J$ to about $1.55J$ is of the same sign as the quantum renormalization of magnon energies in collinear antiferromagnets and also reflects the multimagnon weight contained in the CLS channel; the ED moments of the $N=20$ cluster, $1.35J$--$2.58J$ depending on the spin channel, show the same trend.

The same phenomenology appears in related ferrimagnets. SSE--SAC calculations for the uniform Lieb-lattice Heisenberg model, the simplest lattice with a $1{:}2$ sublattice imbalance, find a broad, weakly dispersive band near $\omega\approx J$ above the ferrimagnetic magnon branch~\cite{chang2024}, and trimer chains and trimerized lattices exhibit composite excitations whose bandwidths are set by intercluster coupling~\cite{cheng2022,chang2024}. Together with the present results, these findings suggest that a broad, weakly dispersive spectral sector is a generic dynamical signature of imbalance-driven flat bands in unfrustrated quantum magnets, whereas a sharp flat magnon requires extra protection, such as a polarized background or a decoupled cluster limit. The unequal-exchange extension studied here interpolates toward the latter, because for $J_B\to0$ the internal excitations of the four-spin stars become dispersionless.

This behavior distinguishes the present mechanism from kagome antiferromagnetism. In the kagome lattice, flat modes originate from frustrated triangular constraints and accompany a macroscopically degenerate classical ground-state manifold~\cite{harris1992,maksymenko2017}. In the edge-centered honeycomb lattice, by contrast, all nearest-neighbor bonds can be satisfied simultaneously, a Marshall sign convention exists, and the sublattice imbalance enforces ferrimagnetism. The Goldstone mode is quadratic rather than linear, and the localized harmonic branch lies at finite frequency rather than at the bottom of a zero-energy manifold. The hexagonal compact pattern shared by both lattices is therefore of geometric origin and occurs in two distinct magnetic phases.

The classical Ising model serves as an exact benchmark of the lattice rather than as a model for the finite-temperature behavior of the quantum magnet. Both the decoration transformation and the Wolff simulations reproduce the two-dimensional Ising transition and its critical exponents. The isotropic Heisenberg model with short-range interactions, in contrast, cannot spontaneously break its continuous spin symmetry at finite temperature~\cite{mermin1966}, so that its ordering transition occurs at $T=0$. The quadratic Goldstone mode nonetheless governs a wide low-temperature regime. For $\omega=Dq^2$ with $D=JSa^2/4$, LSWT gives $u_{\rm th}/(NJ)=[\pi\sqrt3/(60S)]t^2+o(t^2)$ and $C/(Nk_B)=[\pi\sqrt3/(30S)]t+o(t)$, where $t=k_BT/J$. These crossover laws hold below the temperature at which the magnon description breaks down, and the infrared divergence of the magnon number makes interaction corrections essential.

The unequal-exchange extension shows how the uniform ferrimagnet emerges from local building blocks. For $J_B=0$, each four-spin star has a spin-1 triplet ground state while the second hub spin is free; a weak coupling $J_B$ generates an alternating spin-1--spin-$1/2$ honeycomb model with $J_{\rm eff}=5J_B/12$. The exact star thermodynamics predicts the intermediate-temperature entropy $\ln6/5$ and susceptibility $T\chi=11/60$, together with the leading correction $-J_B/(12T)$. The simulations show that strong-bond correlations emerge well before the weaker interstar correlations. For every connected cluster with $J_A,J_B>0$, the Lieb--Mattis spin remains $S_0=L^2/2$, so that varying the bond ratio alone does not produce a singlet ground state.

A different, exact flat mode exists in the fully polarized phase. The one-magnon graph Laplacian has a flat eigenvalue $2$ and a maximum eigenvalue $5$, which yields $\epsilon_{\rm flat}/J=h/J-2S$ and $h_{\rm sat}/J=5S$. This flat band is protected by the one-magnon reduction and is distinct from the zero-field continuum obtained from SSE and SAC.

Photonic super-honeycomb lattices already realize the five-site connectivity and its compact single-particle states~\cite{zhong2017,yan2020}, and an optical-lattice scheme for the same geometry has been proposed~\cite{lan2012}. The recent observation of imbalance-driven ferrimagnetism with ultracold fermions in an optical Lieb lattice~\cite{lebrat2026} shows that the static part of the present physics can be engineered in quantum simulators; a realization of the spin model on the edge-centered honeycomb lattice requires, in addition, a Mott-insulating regime with dominant and nearly uniform antiferromagnetic nearest-neighbor exchange. Among solid-state systems, known candidates instead involve kagome frustration, unequal spin lengths, anisotropy, interlayer coupling, or multiorbital electronic structure~\cite{misumi2020,pratt2025,yoruk2025,lobocheca2024,wang2023,frezza2024}. In a material realization, inelastic neutron scattering would probe the predicted spectrum directly: the quadratic Goldstone mode at low energy and, above it, a broad finite-frequency band whose weak dispersion and edge-sublattice character distinguish it from the dispersive magnon branches. The model therefore establishes a design principle: sublattice imbalance combines robust ferrimagnetism with localized spectral weight, while interactions determine whether this weight forms a sharp mode or a continuum.

\section{Conclusion}
\label{sec:conclusion}

We have shown that sublattice imbalance provides a unifying principle for the edge-centered honeycomb antiferromagnet. The $2:3$ imbalance enforces a single-particle flat band, fixes an extensive ferrimagnetic ground-state spin, and produces an edge-localized flat branch in linear spin-wave theory. Unbiased quantum Monte Carlo simulations establish ferrimagnetic long-range order in the ground state. In the interacting spectrum, the localized structure survives as a weakly dispersive finite-frequency continuum rather than as a protected sharp magnon. This transformation of a harmonic flat band into a structured many-body continuum distinguishes imbalance-driven localization from frustration-induced flat modes in kagome magnets and identifies a route to ferrimagnets in which robust static order coexists with localized many-body spectral weight.

\section*{Data availability}

The data supporting the conclusions of this investigation and the numerical implementation are available from the corresponding author upon reasonable request.

\section*{Acknowledgments}
This work is funded by Natural Science Foundation of Hebei Province (Grant No. A2026203033), Beijing National Laboratory for Condensed Matter Physics (Grant No. 2024BNLCMPKF020) and Innovation Capability Improvement Project of Hebei province (Grant No. 22567605H).  

\bibliography{references_prb_verified}

\end{document}